\documentclass[journal]{IEEEtran}
\usepackage{graphicx}
\usepackage{amsmath}
\usepackage{subfigure}
\usepackage{multirow}
\usepackage{array}
\ifCLASSINFOpdf

\else

\fi

\begin{document}
\title{Modeling Enhancements in DSR, FSR, OLSR under Mobility and Scalability Constraints in VANETs}

\author{\IEEEauthorblockN{N. Javaid, A. Bibi, S. H. Bouk, A. Javaid, I. Sasase$^{\$}$\\}%\vspace{0.4cm}}
%    \IEEEauthorblockA{ \{nadeem.javaid,djouani@univ-paris12.fr\}}\\
        Department of Electrical Engineering, COMSATS, Islamabad, Pakistan. \\
        $^{\$}$Department of Information and Computer Science, Keio University, Japan.

     }
\vspace{-2cm}

%
%% make the title area

\maketitle

\begin{abstract}
%\boldmath
Frequent topological changes due to high mobility is one of the main issues in Vehicular Ad-hoc NETworks (VANETs). In this paper, we model transmission probabilities of $802.11p$ for VANETs and effect of these probabilities on average transmission time. To evaluate the effect of these probabilities of VANETs in routing protocols, we select Dynamic Source Routing (DSR), Fish-eye State Routing (FSR) and Optimized Link State Routing (OLSR). Framework of these
protocols with respect to their packet cost is also presented in this work. A novel contribution of this work is enhancement of chosen protocols to obtain efficient behavior. Extensive simulation work is done to prove and compare the efficiency in terms of high throughput of enhanced versions with default versions of protocols in NS-2. For this comparison, we choose three performance metrics; throughput, End-to-End Delay (E2ED) and Normalized Routing Load (NRL) in different mobilities and scalabilities. Finally, we deduce that enhanced DSR (DSR-mod) outperforms other protocols by achieving $16\%$ more packet delivery for all scalabilities and $28\%$ more throughput in selected mobilities than original version of DSR (DSR-orig).
\end{abstract}


\IEEEpeerreviewmaketitle

%section1
\section{Introduction}

Wireless networks offer convenience to the users in different nodes' mobilities and densities. Vehicular Ad-hoc NETworks (VANETs) are distributed and self-assembling communication networks made up of multiple autonomous moving vehicles, and are associated with high mobilities. Their major purpose is to provide safety and ease to the travelers. Vehicles are equipped with VANETs' device which can become a node in ad-hoc network and can search out and pass on messages through the wireless network among the nodes.

In wireless networks, the routing protocols which calculate efficient routes for end-to-end connectivity, are of two types; reactive and proactive. Protocols belong to former type calculate routes for destination in network when data demands arrive, whereas, in later type, routes are calculated periodically independent from data demands. Routing overhead in terms of routing load and path latencies is a critical issue to be tackled for achieving high delivery rates. Routing protocols, in this context, are aimed to optimize these issues.

To study routing protocols in VANETs, we select three protocols; a reactive protocol Dynamic Source Route (DSR) [1], and two proactive protocols; Fish-eye State Routing (FSR) [2] and Optimized Link State Routing (OLSR) [3]. After analyzing these protocols, we enhance them and compare their efficiency in Nakagami radio propagation model for VANETs using NS-2.

%section2
\section{Background}

Authors in [4] compare and evaluate performance of Ad-hoc On-demand Distance Vector (AODV), DSR, and Swarm Intelligence based protocol. They perform a variety of simulations for VANETs, characterized by networks' mobility and size.

[5] evaluates AODV and OLSR in realistic urban scenarios and studies the chosen protocols under different metrics such as vehicles' mobility, density and data traffic rates.

DYnamic MANET On-demand (DYMO) routing protocol is analyzed by the authors in [6]. In order to evaluate the performance of typical ad hoc routing protocols, they combine micro simulation of road traffic and event-driven network simulation. In their work, using different parameters of DYMO for a multitude of traffic and communication scenarios improve the overall performance.

A comprehensive evaluation of mobility impact on IEEE $802.11p$ MAC performance is performed in [7]. This study evaluates packet delivery ratio, throughput, and delay as performance metrics. Authors also propose two dynamic contention window mechanisms to alleviate network performance degradation due to high mobility. Their extensive simulation results demonstrate a significant impact of mobility on the IEEE $802.11p$ MAC performance.

\section{Motivation}

AODV and OLSR is evaluated in urban scenarios by Khan. I \textit{et. al} in [8]. They enhance HELLO and TC interval of OLSR and observe that overall enhanced OLSR perform better than AODV in urban environments. They select Packet Delivery Ratio (PDR), E2ED and Routing Packets per Data Packets (NRL in other words) as performance metrics for evaluating different vehicles' scalabilities using probabilistic Nakagami radio propagation model in NS-2.

In our paper, we construct a model for: (1) probability of transmission in $802.11p$ at MAC layer, (2) energy cost of routing protocols at network layer. We enhance OLSR in the same way as that in [8]. Moreover, we also enhance DSR and FSR as well. Throughput, E2ED and NRL performance metrics are selected for evaluating performance of routing protocols in VANETs. For mobility analysis $0$, $100$, $200$ and $400\,\, Pause\,Time\,(s)$ with speed of $30m/s$, and $10$, $20$, $30$ and $40\,no.\,\,of\,\,connections$ for scalability analysis are selected using probabilistic Nakagami radio propagation model NS-2.

\section{System Model}

In [7], the back-off procedure of IEEE $802.11p$ is p-persistent CSMA/CA is used to offer adaptivity for neighbor nodes. The major variation between p-persistent $802.11$ and standard IEEE $802.11p$ protocol is only based on assortment of back-off interval. In standard protocol, back-off interval is binary exponential. However, in p-persistent CSMA/CA, the back-off interval is supported on a geometric distribution with a transmission probability, $p$. Therefore, the probability that a node stays idle when having a busy medium is $1-p$. Based on the geometrically dispersed back-off time, the probability of having a successful transmission after $n-1$ failures is:

%eq1
\begin{eqnarray}
P(X=Z)=(1-p)^{Z-1}p,\,\,\,\,\,Z=1,2,3,.....
\end{eqnarray}

Where, $X$ is a number represents the total trials for a successful transmission in a contention window, $CW$. Based on [9], [10], the predictable value of X can be used to determine the average Contention Window size $\overline{CW}$ as:

%eq2
\begin{eqnarray}
E[X]=\sum_{Z=1}^{\infty}Z_p(1-p)^{n-1}=\frac{1}{p}
\end{eqnarray}

%eq3
\begin{eqnarray}
\frac{\overline{CW}+1}{2}=\frac{1}{p}\frac{}{}
\end{eqnarray}

Let $Q$ is the number of contending nodes, then there are several probabilities. Let the probability of a successful transmission, $P_s$, and the probability of a collision, $P_c$, are $P_s=P\{One \,node\, tranmists\,/At\, least\, one \,node \,has \,a \,packet \,to\\ \,transmit\}$ and $P_c=P\{Two \,nodes \,tranmist\,/AT \,least \,one\\ \,node \,has \,a \,packet \,to \,transmit\}$, respectively. These probabilities are given below:

%eq4,5,6
\begin{eqnarray}
P\{No\,\,transmission\}=(1-p)^{Q}\\
P\{Only\,\,one\,\,transmission\}=Qp(1-p)^{Q-1}\\
P\{At\,\,least\,\,one\,\,transmission\}=1-(1-p)^{Q}
\end{eqnarray}

%eq7
\begin{eqnarray}
P_s=\frac{Qp(1-p)^{Q}}{1-(1-p)^Q}
\end{eqnarray}

%eq8
\begin{eqnarray}
P_c=\frac{1-(1-p)^{Q}-Qp(1-p)^{Q-1}}{1-(1-p)^Q}
\end{eqnarray}
\normalsize

The time interval between two adjacent successful transmissions is defined in [9] as a virtual transmission time; $\tau _{v-trans}$. It is possible to have a number of collisions in addition to one successful transmission, in a $E[\tau _{v-trans}]$. Let $\tau _{idle}$ denotes the idle time during which no node transmits, $\tau _{succ}$  the time of a successful transmission, and $\tau _{coll}$ the total time of transmission collisions, within an average virtual transmission time $E[\tau _{v-trans}]$, then from [9], we have:

%eq9
\begin{eqnarray}
E[\tau _{v-trans}]=E[\tau _{idle}]+E[\tau _{coll}]+E[\tau _{succ}]
\end{eqnarray}

Where,

%eq10
\begin{eqnarray}\nonumber
E[\tau _{idle}]=\frac{1-(1-p)^{Q}-Qp(1-p)^{Q}}{Qp(1-p)^{Q-1}}\times\frac{1-p}{Qp}\tau _{slot},
\end{eqnarray}

%eq11
\small
\begin{eqnarray}\nonumber
E[\tau _{coll}]=\frac{1-(1-p)^{Q}-Q_p(1-p)^{Q}}{Qp(1-p)^{Q-1}}(\tau _{pack}+\tau _{DIFS})\tau _{slot}, \,\,\, and
\end{eqnarray}
\normalsize

%eq12
\begin{eqnarray}\nonumber
E[\tau _{succ}]=(\tau _{pack}+\tau _{DIFS})\tau _{slot}.
\end{eqnarray}

The value of  $\tau _{v-trans}$ should be minimized for maximum system efficiency in terms of throughput. Let $\tau _{pack}$, $\tau _{DIFS}$ and $\tau _{slot}$ denote packet transmission time, DIFS time, and slot time, respectively. Based on the probabilities of transmissions and a constant packet time $\tau _{pack}$, mathematical expressions for $\tau _{idle}$, $\tau _{succ}$ and $\tau _{coll}$ from [7] and [9] as:

%eq10
\small
\begin{eqnarray}
E[\tau _{v-trans}]=\frac{(\tau _{pack}+\tau _{DIFS})-(\tau _{pack}+\tau _{DIFS}-1)(1-p)}{Qp(1-p)^{Q-1}}\tau _{slot}
\end{eqnarray}
\normalsize

%section3
\section{Routing Operations in DSR, FSR and OLSR}

The detailed description of selected protocols with their modeled energy cost is given below:

\subsection{DSR}

Reactive protocols are on-demand in nature and start route finding for requested destination when data request arrives. This process is known as Route Discovery (RD) and permits any host to randomly discover a route for any other host in the network that is either directly reachable or by relay nodes.

In DSR, before broadcasting Route REQuest (RREQ), the originator checks route for the requested target in Route Cache (RC), if path for target is not present then broadcast RREQ. This mechanism of searching routes in RC during RD is known as RCing and is possible due to storage of learned routes due to promiscuous listening mode. RD is performed by Expanding Ring Search (ERS) mechanism, the Energy Cost of RD;  $C_{E-RD}^{(DSR)}$ from [11]:

%eq11
\small
\begin{eqnarray}
 C_{E-RD}^{(DSR)}=
  \begin{cases}
   \displaystyle\displaystyle\sum_{R_i=1}^{R_{max\_limit}}(C_{E-R_i})_{R_i} & \hspace{-1cm} if\, no\, RREP\, received \\
   C_{E-R_{rrep}} & \hspace{-1cm} if\, TTL(R_{rrep})=1 \\
   {\displaystyle\displaystyle\sum_{R_i=1}^{R_{rrep}}(C_{E-R_i}})_{R_i} & \hspace{-1cm} otherwise\\
   %\hspace{0.4in}
   \{R_{rrep}=1,2,3,....,max\_limit\}
  \end{cases}
\end{eqnarray}
\normalsize

DSR uses ERS mechanism and thus broadcasts RREQ through different rings, $R_1,R_2,R_3,...,R_{rrep}$, where $R_{rrep}$ generates RREPs and further broadcasting is stopped (Fig. 1).

In reactive protocols after establishment of successful routes Route Maintenance (RM) process is started. This process involves Link Status Monitoring ($LSM$) and Route Repairing (RR) phases. $LSM$ is used to check the connectivity of active routes which are established successfully during RD. If link breakage is reported during $LSM$, then the next task is to repair the route. This repairing process is performed during RR phase. This phase involves dissemination of Route ERRor (RERR) message about broken link and route re-discovery for broken route.

After detecting link breakage, in DSR, the node which detects link breakage first search RC for alternative route. If it finds alternative route then sends the data to this route, otherwise, searching alternative route in RC is repeated by each node in active route. This repairing process is known as Packet Salvaging (PSing). In case of unsuccessful PSing, source node starts new RREQ  route re-discovery process based on $MaxMaintRexmt$ constraint [1]. Moreover, RERR message is piggy-backed with new RREQ.

\begin{figure}[h]
\begin{center}
\includegraphics[
height=5 cm,
width=7 cm
]{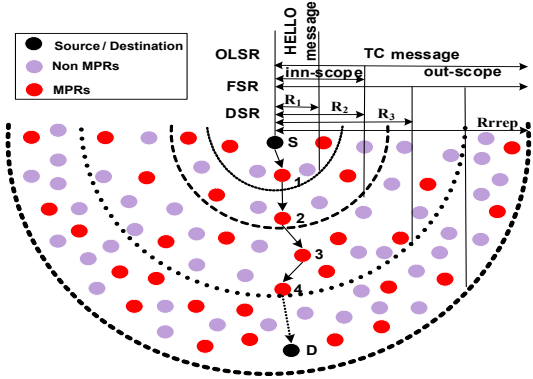}
\end{center}
\vspace{-0.5cm}
\caption{Route Calculation in DSR, FSR, and OLSR}
\end{figure}

\subsection{FSR}

Proactive routing protocols perform three different operations for maintaining network topology and routes information in routing table; Link Status Monitoring Periodically $(LSM\_Per)$, Route Updates Triggered $(RU\_Tri)$ and Routes Updates Periodic $(RU\_Per)$. $LSM\_Per$  maintains recent information about Link Status ($LS$) in the network and check the connectivity of a node in the network.  $RU\_Per$ operation updates routing information across the network. Whenever $LS$ changes then $RU\_Tri$'s are generated.

FSR implements Scope Routing (SR) and generates only periodic updates; $LSM\_Per \,and\,RU\_Per$. To exchange routing information in SR technique, $RU_{Per}$ are disseminated in different scopes based on search diameter in terms of number of hops. Graded-frequency technique is associated with each scope to reduce routing packets' overhead. Two scope levels are used for FSR in [2], one is known as IntraScope and second is known as InterScope. Energy cost for former scope as $C_E^{(in)}$ with search diameter of $2$ hops, and energy cost for later scope as $C_E^{out-sco}$ with number of hops value of $255$, respectively is portrayed in Fig. 1. From [12], per packet energy cost for FSR $(C_E^{(FSR)})$ is:

%eq12
\begin{eqnarray}
\begin{split}
\hspace{-1cm}\displaystyle C_E^{(FSR)}=& \int _{0}^{\tau}d_{avg}^{out}\sum_{i=1}^{N_{out}-1} (p_{err})^{i+1}\prod _{j=1}^{i}d_f[j] \\ 
&+ d_{avg}^{in}\sum_{i=1}^{N_{in}-1} (p_{err})^{i+1}\prod _{j=1}^{i}d_f[j]
\end{split}
\end{eqnarray}
\normalsize

\subsection{OLSR}

OLSR performs only $RU\_Tri$ in the entire network for maintaining fresh routes, while sends $LSM\_Per$ through HELLO messages at routing layer. Let $C_{E}^{(OLSR)}$ denotes total energy cost of OLSR.

\vspace{-0.3cm}
%eq13
\begin{eqnarray}
C_{E}^{(OLSR)}=\int _{0}^{\tau_{LU}} C_{E-nc}^{MPR} + C_{E-c}^{MPR}
\end{eqnarray}

Where, $\tau_{LU}$ is the last update message generated during network life time. The interval for transmission of routing updates varies with respect to the status of MPR. If MPRs' status remains the same, then TC messages are transmitted through default TC interval (Table. 1). Whereas, TC messages are triggered whenever MPRs' status changes. $C_{E-nc}^{MPR}$ is the cost of allowed (re)transmissions through MPRs, while $C_{E-c}^{MPR}$ shows the cost of dissemination of update messages in the whole network [12].

\vspace{-0.3cm}
%eq14
\small
\begin{eqnarray}
C_{E-nc}^{MPR}=(1-p_c^{MPR})p_{err}d_{avg}+d_{avg}\sum_{i=1}^{h-1} (p_{err})^{i+1}\prod _{j=1}^{i}d_f^{MPR}[j]
\end{eqnarray}
\normalsize

\vspace{-0.3cm}
%eq15
\small
\begin{eqnarray}
C_{E-c}^{MPR}=p_c^{MPR} p_{err}d_{avg}+d_{avg}\sum_{i=1}^{h-1} (p_{err})^{i+1}\prod _{j=1}^{i}d_f[j]
\end{eqnarray}
\normalsize

Fig.1 shows OLSR's HELLO messages are exchanged with their neighbors, and TC messages' transmission in entire network only through MPRs.

     \begin{table}[!h]
    \centering
    \small
    \begin{tabular}{|m{1cm}|m{4cm}|m{1cm}|m{1cm}|}
    \multicolumn{4}{c}{Table.2. Default and Modified Parameters of Selected Protocols} \\
    \hline
    \textbf{Protocol}&\textbf{Parameter}&\textbf{Default Value}&\textbf{Modified Value}\\
    \hline
    \textbf{DSR}&$NonPropagating RREQ$&1&3\\\cline{2-4}
    &$TAP\_CACHE\_SIZE$&1024&256\\\hline
    \textbf{FSR}&$IntraScope\_Interval$&5s&1s\\\cline{2-4}
    &$IntreScope\_Interval$&15s&3s\\\hline
    \textbf{OLSR}&$HELLO\_INTERVAL$&2s&1s\\\cline{2-4}
    &$TC\_INTERVAL$&5s&3s\\\hline
    \end{tabular}
    \normalsize
     \end{table}

%section4
\section{Simulations and Discussions}
In VANETs, delay is a critical issue. To tackle this issue, we make enhancement in selected protocols. In DSR, we have modified $NonPropagating RREQ$ value from $1$ to $3$ and $TAPE\_CACHE\_SIZE$ from $1024$ to $256$. For RD, DSR uses ERS mechanism and initiates RREQ from $NonPropagating RREQ$. In ERS, gradual increase of search diameter and time consumption values depend on previous TTL value and waiting time. Thus, incrementing $NonPropagating RREQ$ results quick search and and minimizes searching delay. Moreover, DSR implements PSing and RCing mechanisms due to premisses listening mode for RD and RM, respectively. Due to absence of explicit mechanism for stale route deletion, faulty routes are stored in $RC$ because of larger value of $TAPE\_CACHE\_SIZE$. Reducing $TAPE\_CACHE\_SIZE$ avoid the storage of faulty routes. Therefor, reduction of $TAPE\_CACHE\_SIZE$ values (Table. 1) results fruitful PSing and RCing, and lessens routing delay. These enhancements work well for high mobile and populated scenarios.

Delay in updates of routing entries causes low convergence in high mobile networks. Being a high mobile network, VANET demands low latencies for better efficiencies in terms of accurate data delivery. To reduce this delay, we have shorten periodic updates' intervals for both FSR and OLSR. In FSR, intervals of both scopes; $IntraScope\_Interval$ and $InterScope\_Interval$ are reduced (Table. 1). Whereas, intervals of $LSM_{Per}$ and $RU_{Tri}$  i.e., $HELLO\_INTERVAL$ and $TC\_INTERVAL$, respectively are shortened in enhanced OLSR (Table. 1).

Performance of the chosen protocols with their default and enhanced versions has been evaluated and compared with three performance parameters; throughput, E2ED, and NRL in NS-2. The detail of simulation parameters is given in Table. 2.

\begin{table}[!h]
\centering
\small
\begin{tabular}{|c|c|}
    \multicolumn{2}{c}{Table.1. Common Simulation Parameters} \\
    \hline
    \textbf{Parameter}&\textbf{Value}\\
    \hline
    MAC Protocol&$802.11p$\\\hline
    Area & 1000 x 1000 $m^2$\\ \hline
    Simulation Time& 900 Seconds \\ \hline
    Data Traffic Source & CBR of 512 bytes \\
    \hline
    Mobility Model&Nakagami Model\\
    \hline
   % Wireless Link Bandwidth&2Mbps\\
%    \hline
    Mobility& 0, 100, 200, 400 Pause Time (s)\\
    \hline
    Scalability& 25, 50, 75, 100 nodes\\\hline
\end{tabular}
\normalsize
\end{table}

\begin{figure*}[!t]
\centering
\subfigure[Different Scalabilities]{\includegraphics[height=3.2 cm,width=5.5 cm]{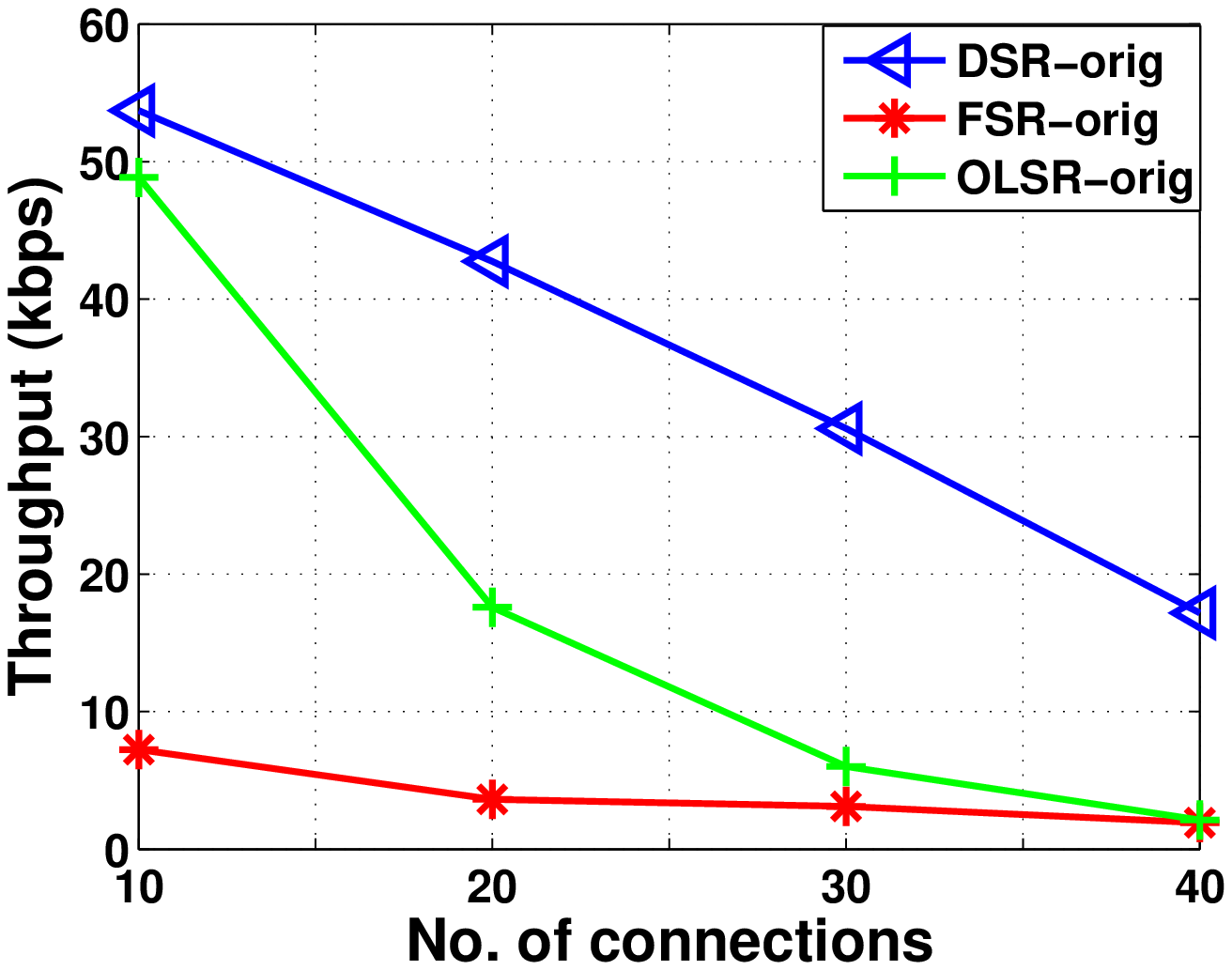}}
\subfigure[Different Scalabilities]{\includegraphics[height=3.2 cm,width=5.5 cm]{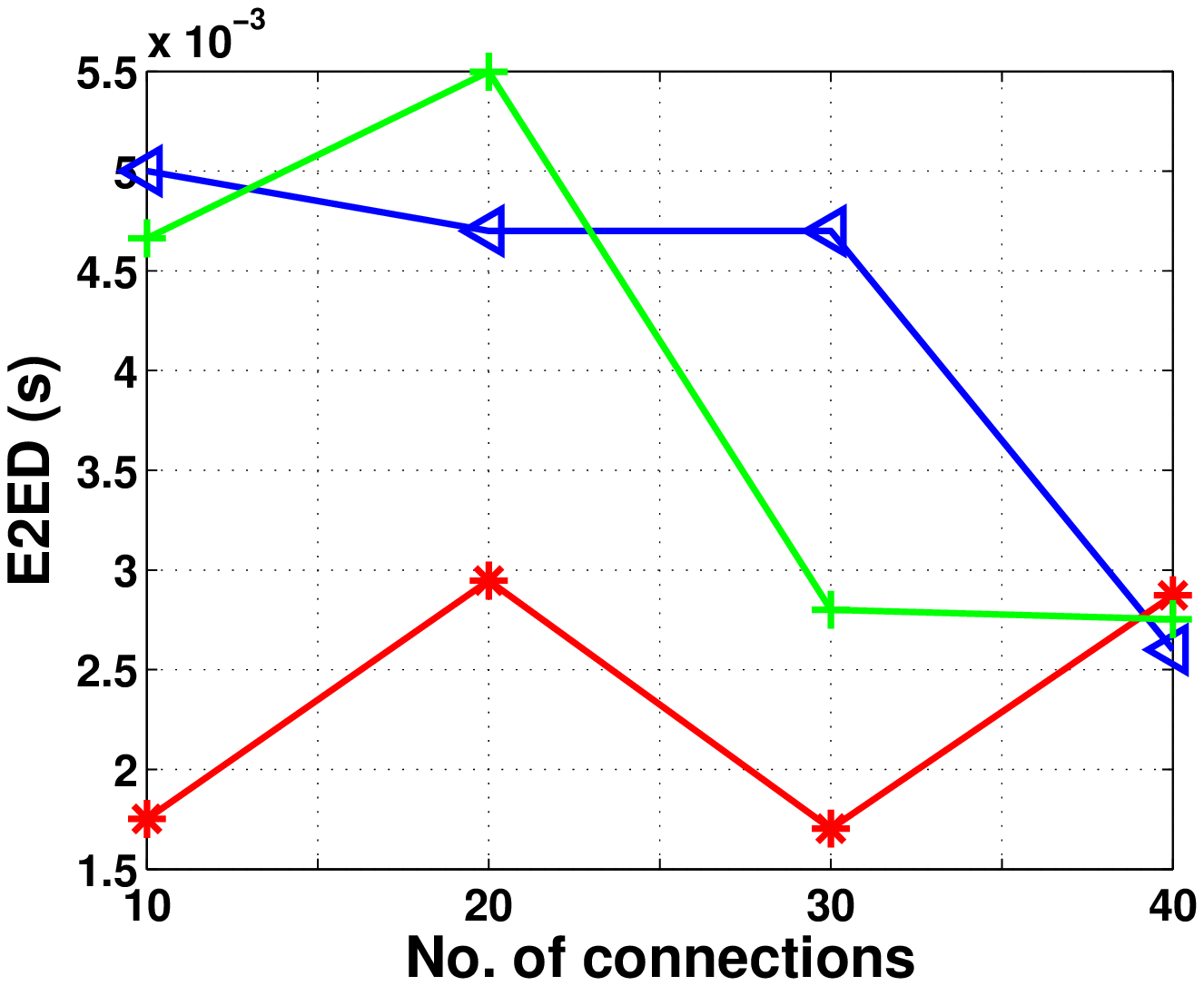}}
\subfigure[Different Scalabilities]{\includegraphics[height=3.2 cm,width=5.5 cm]{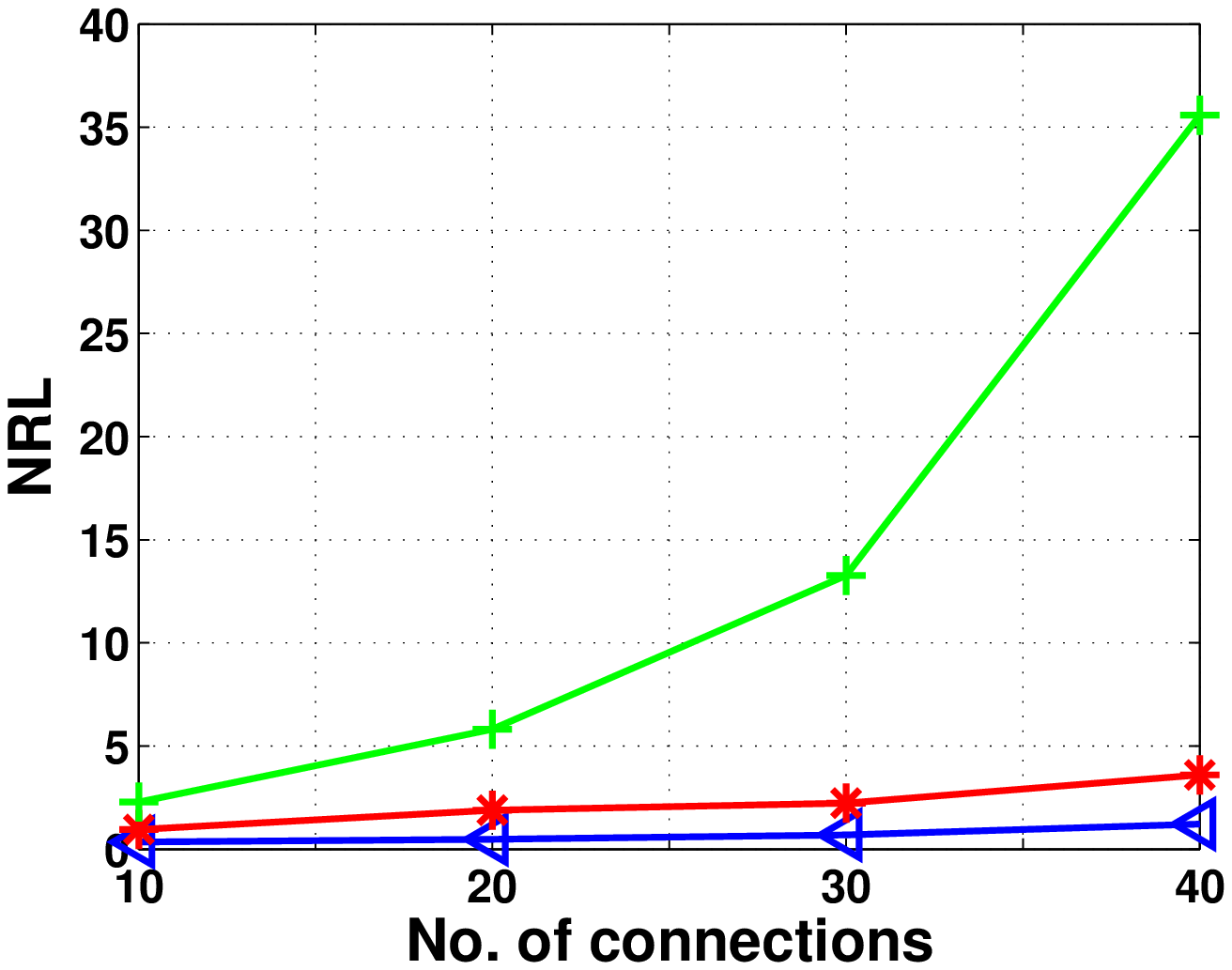}}

\subfigure[Different Scalabilities]{\includegraphics[height=3.2 cm,width=5.5 cm]{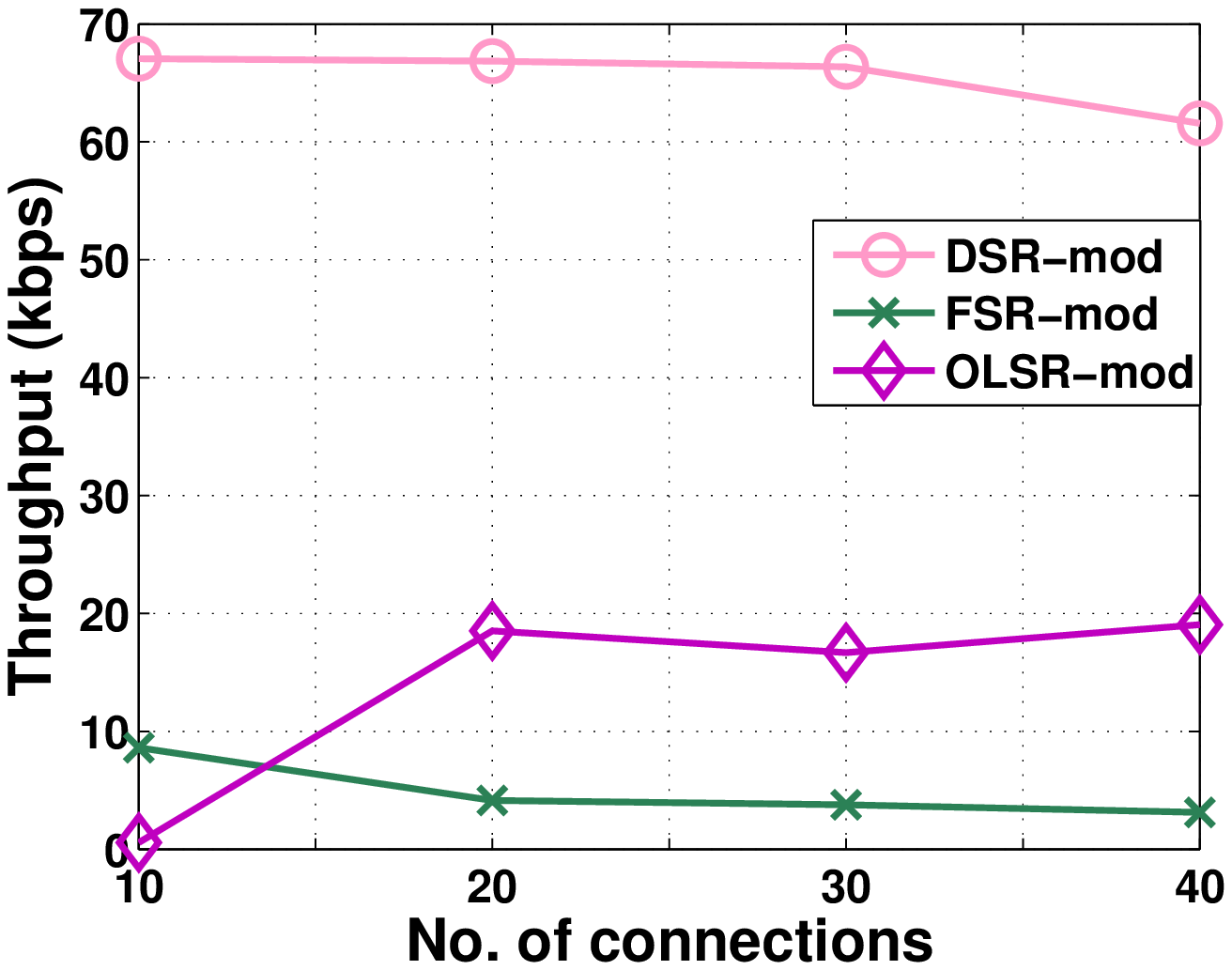}}
\subfigure[Different Scalabilities]{\includegraphics[height=3.2 cm,width=5.5 cm]{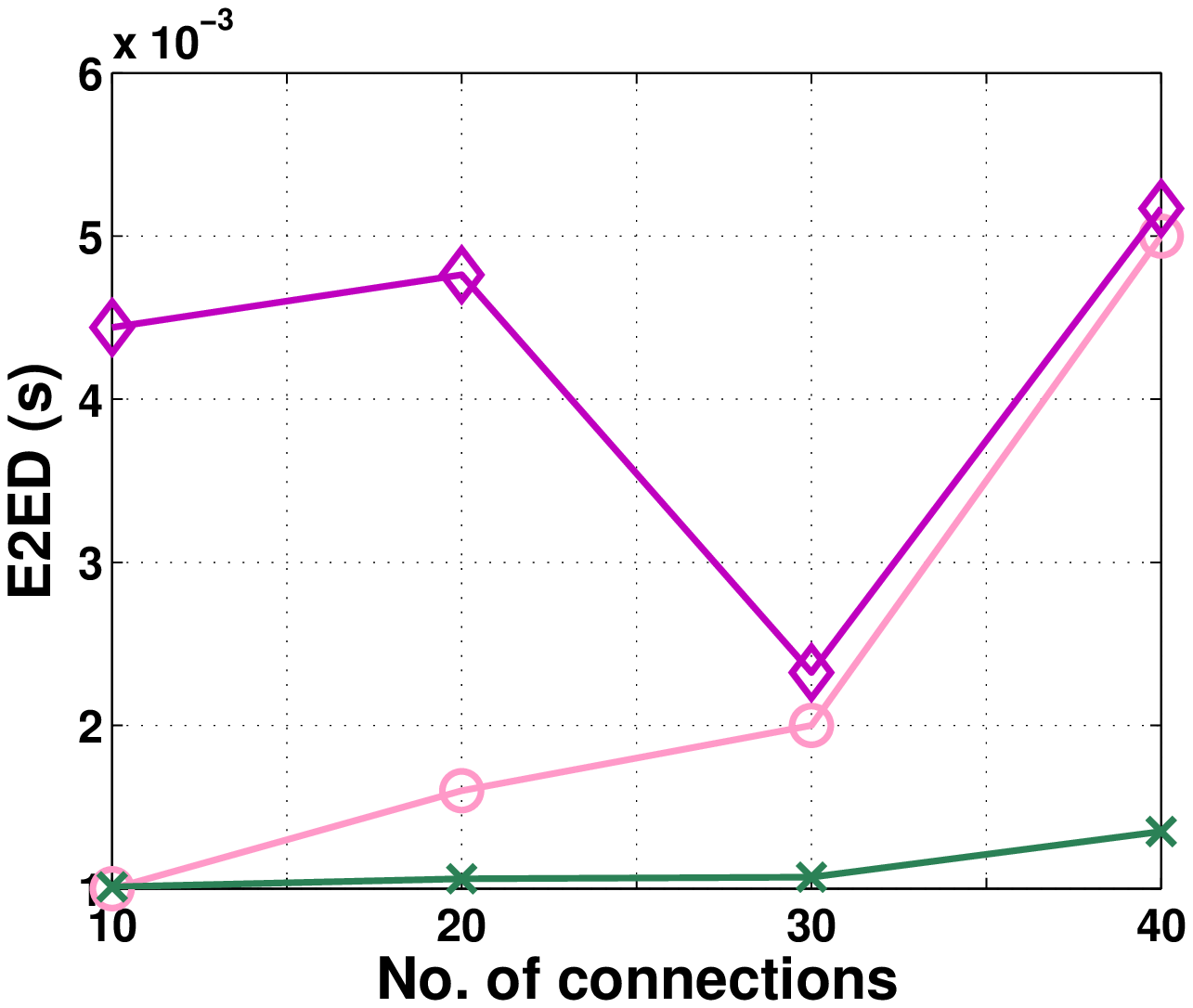}}
\subfigure[Different Scalabilities]{\includegraphics[height=3.2 cm,width=5.5 cm]{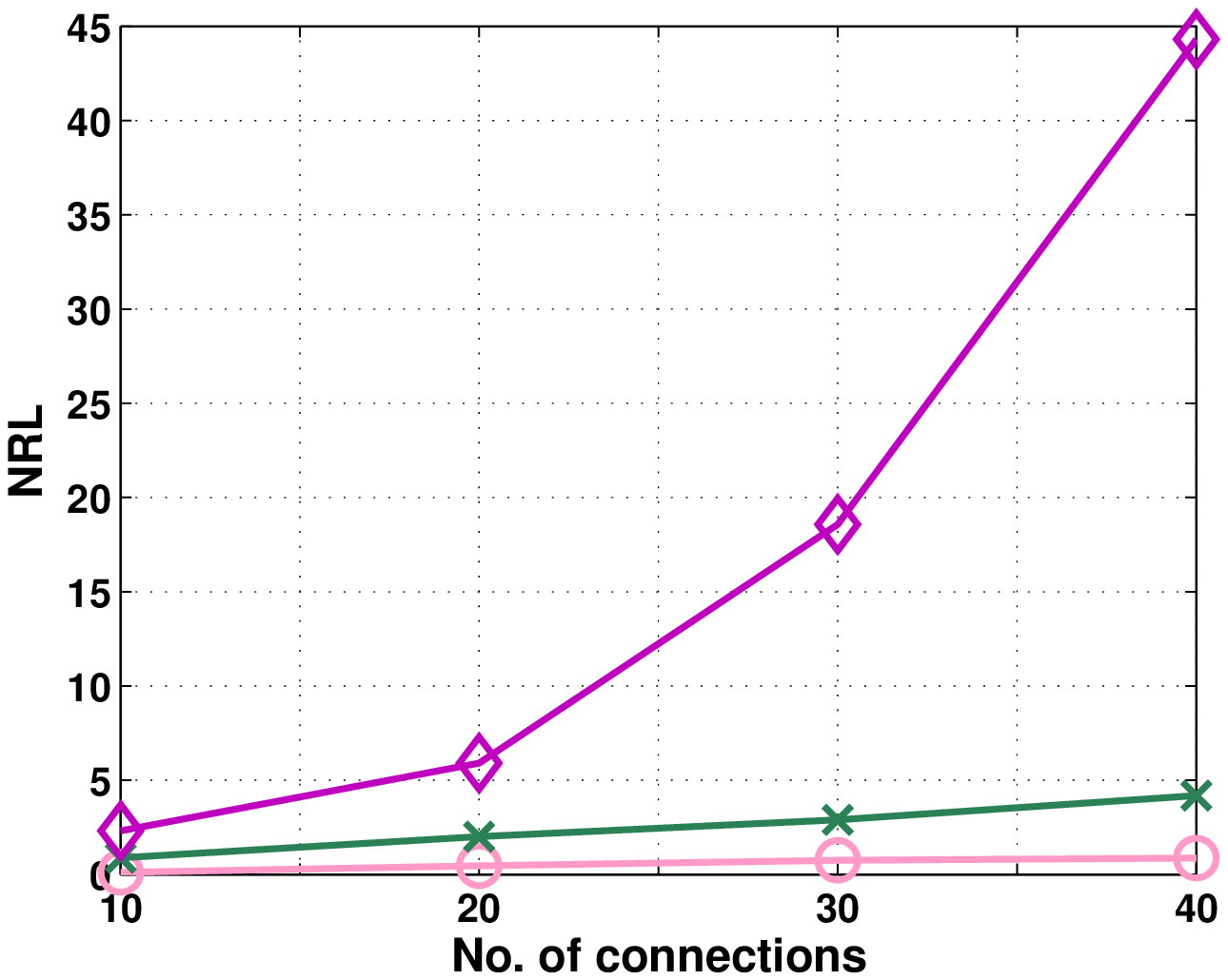}}

\subfigure[Different Mobilities]{\includegraphics[height=3.2 cm,width=5.5 cm]{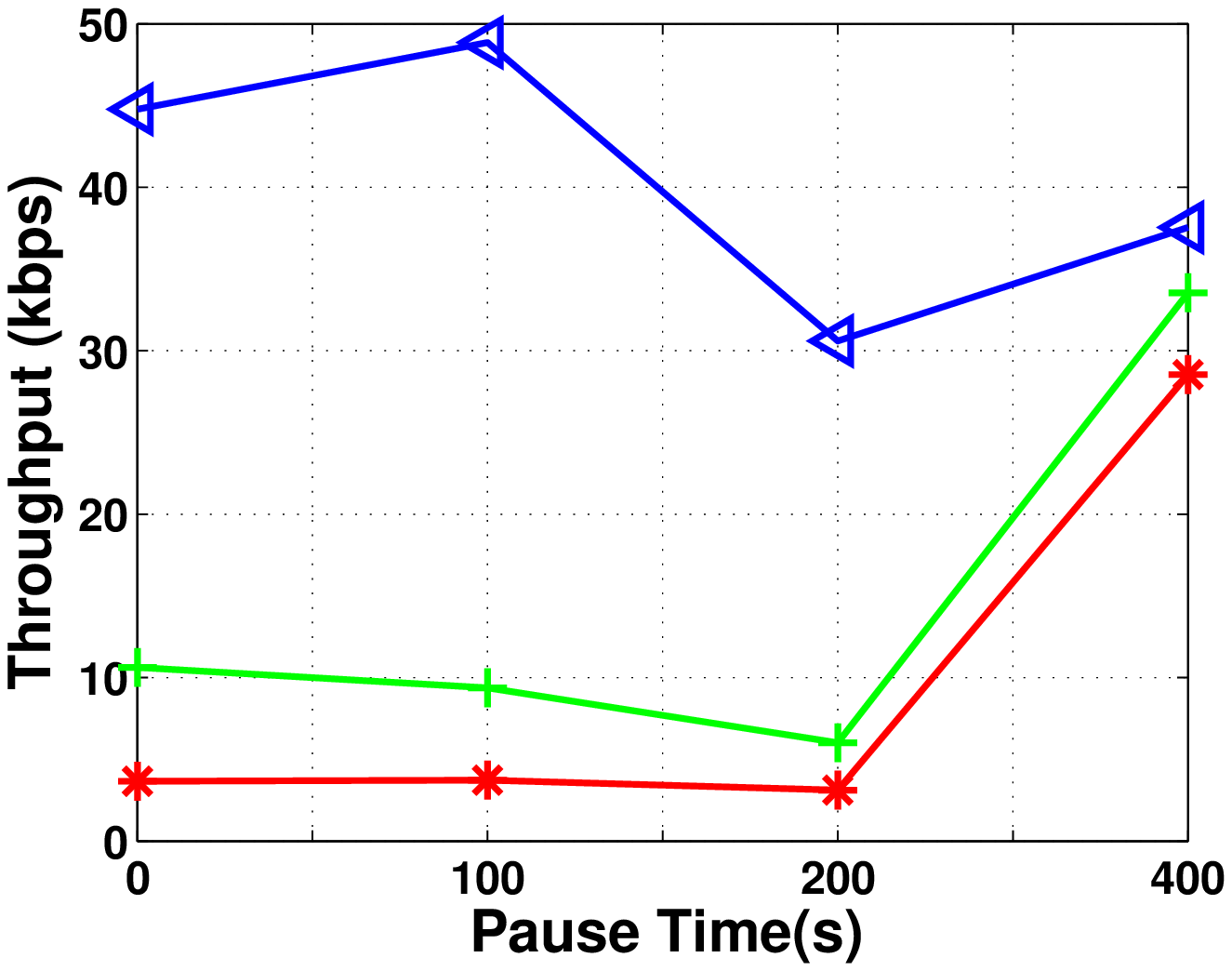}}
\subfigure[Different Mobilities]{\includegraphics[height=3.2 cm,width=5.5 cm]{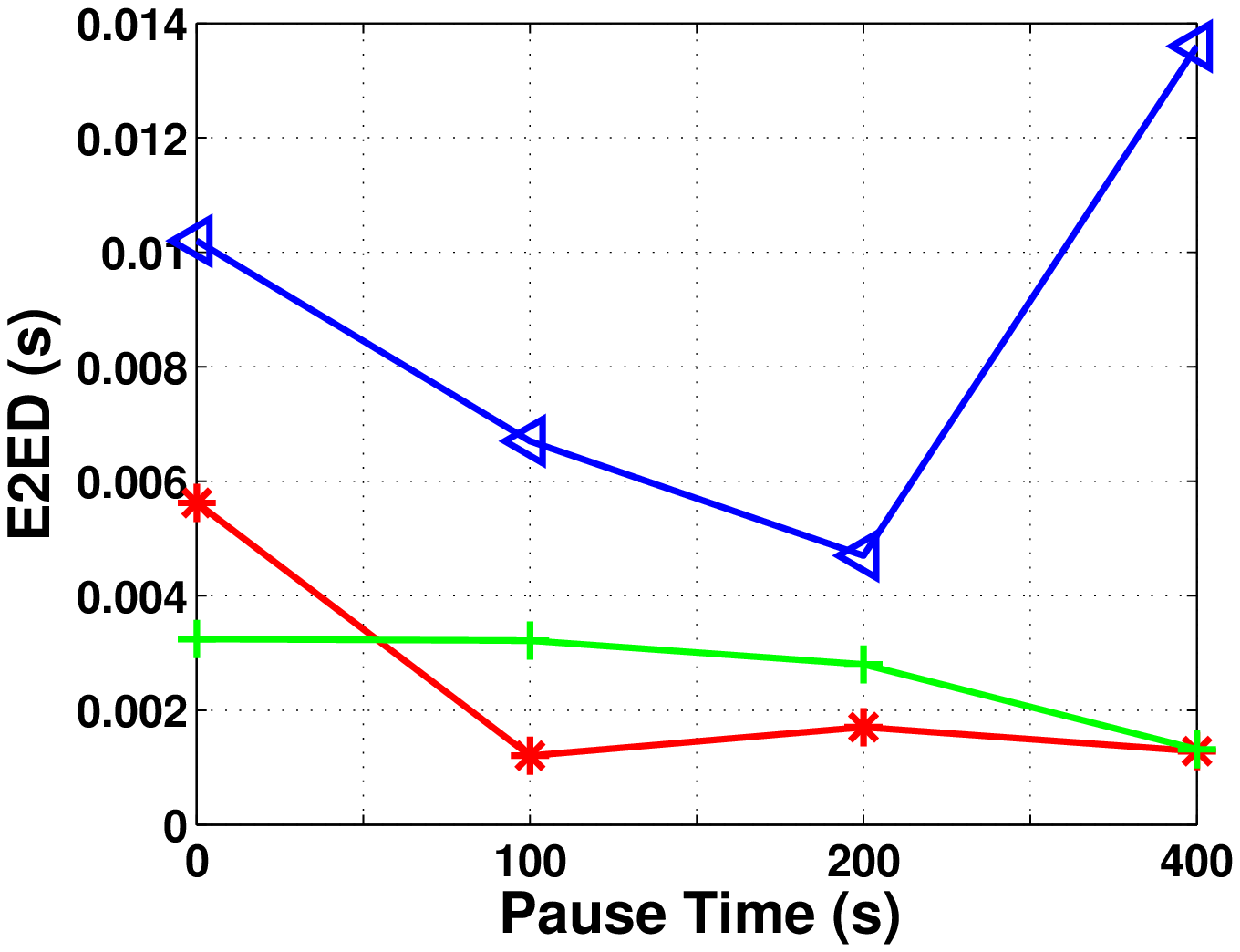}}
\subfigure[Different Mobilities]{\includegraphics[height=3.2 cm,width=5.5 cm]{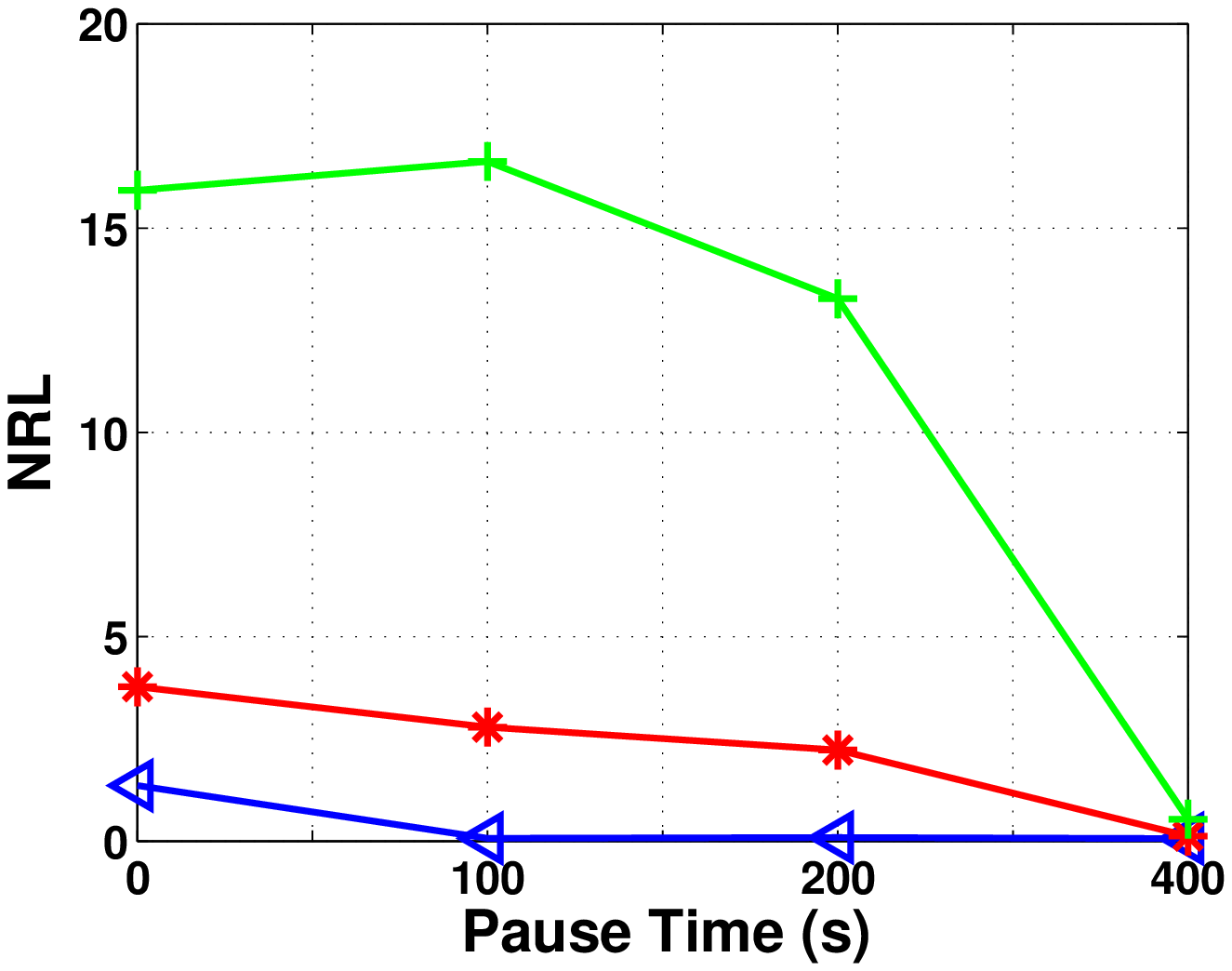}}

\subfigure[Different Mobilities]{\includegraphics[height=3.2 cm,width=5.5 cm]{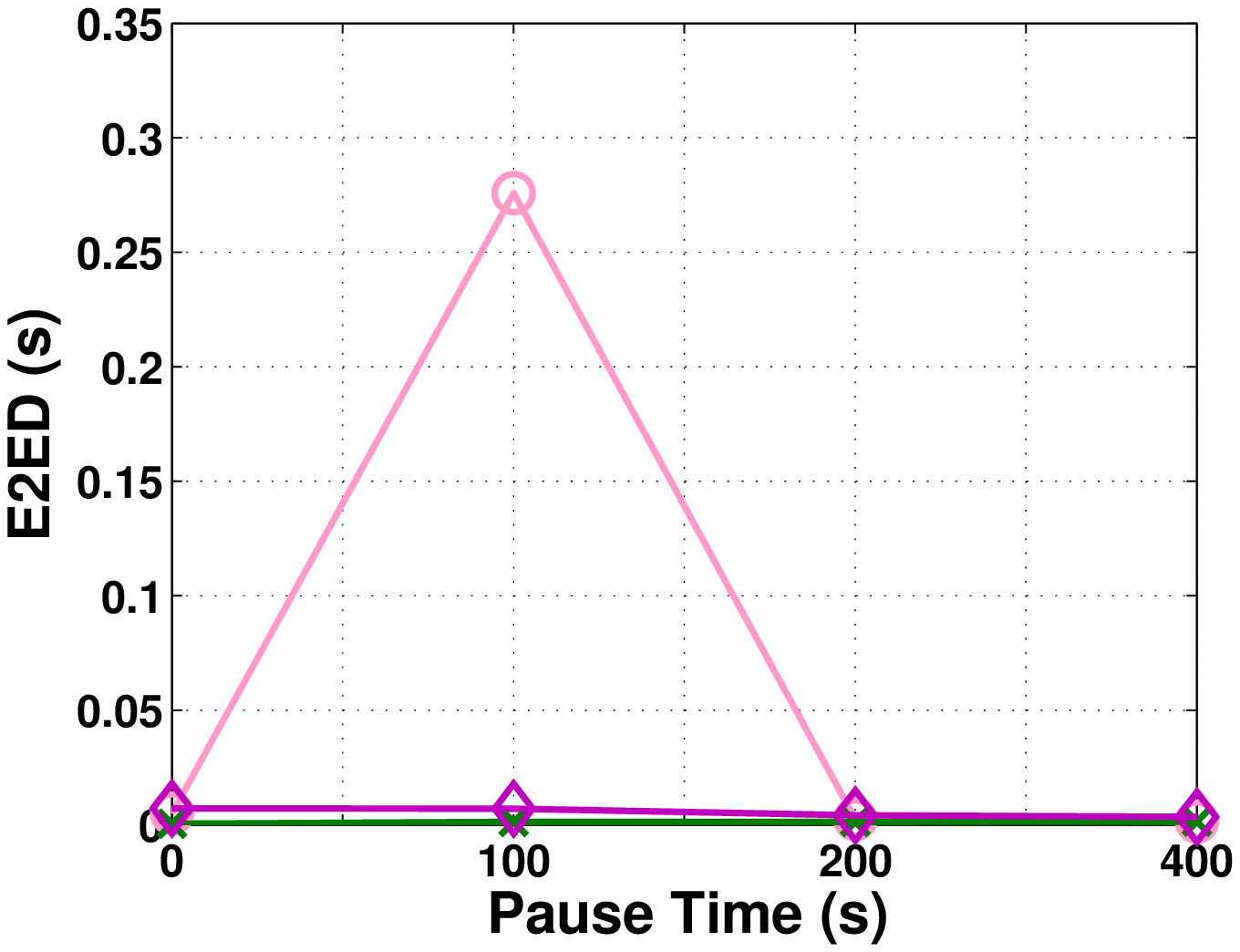}}
\subfigure[Different Mobilities]{\includegraphics[height=3.2 cm,width=5.5 cm]{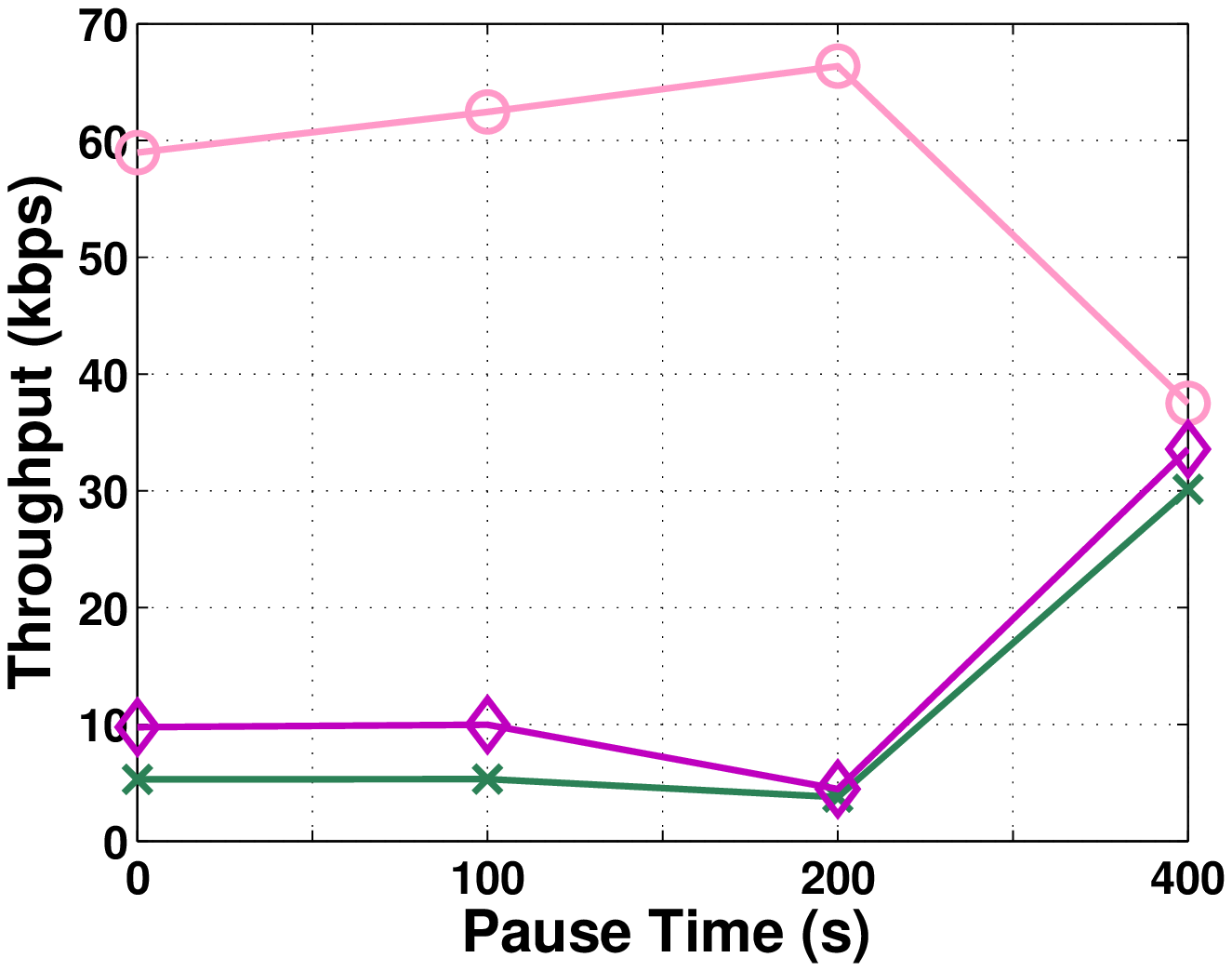}}
\subfigure[Different Mobilities]{\includegraphics[height=3.2 cm,width=5.5 cm]{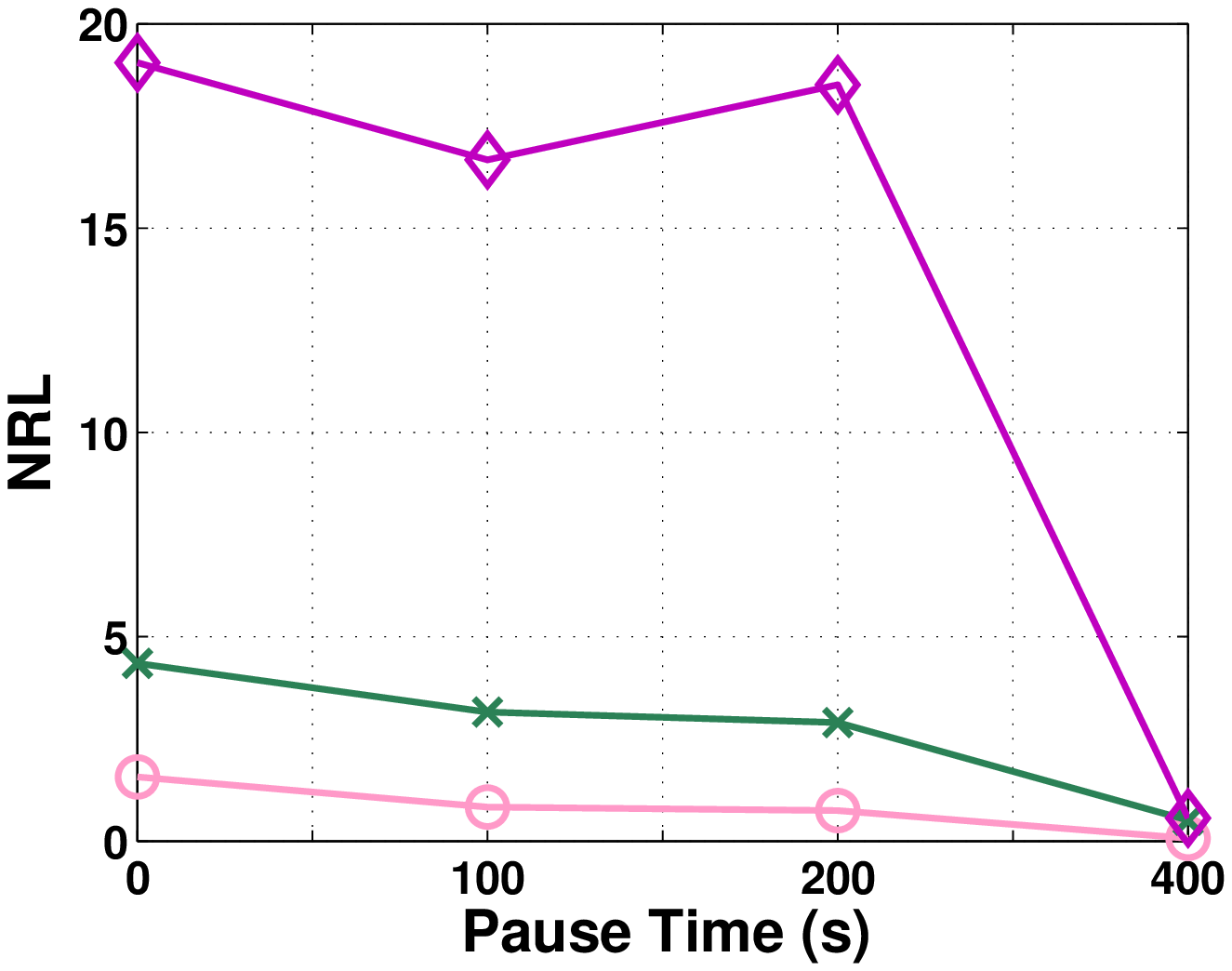}}

\caption{Simulation Results for the Study Conducted in this Work}
\end{figure*}

\subsection{Throughput}

In VANETs, quick convergence is required to exchange data. As, moving vehicles alter the existing established routes therefore, demand quick repairing. DSR among selected protocols produces highest throughput (Fig. 2.a,d,g,j) because of reactive nature. $RU\_Tri$'s are used in OLSR for convergence purpose, whereas, FSR only uses $RU\_Per$, that is why, OLSR's throughput is more than FSR. Throughput of DSR-orig is decreased in case of more no. of connections and in high mobilities as compared to DSR-mod, as shown in Fig. 2.a,d and Fig. 2.g,j. DSR-orig does not scale well in high scalabilities because of generation of grat. RREPs which are produced during RD and RM creating broadcast storm. In DSR-mod, we reduce size of RC by modifying $TAP\_CACHE\_SIZE$ (Table. 1). This change makes the fresh routes available for RCing, thus, remarkable change in throughput value is obtained from Fig. 2.d,j, where $16\%$ and $24\%$ efficiency is achieved by DSR-mod with respect to DSR-orig, in Fig. 2.a,g.

In VANETs, FSR behaves worst among all protocols due to lack of any instance action for link changes (as it uses only periodic operations). After shortening the scope interval in FSR-mod, routing updates are disseminated frequently thus, throughput becomes $6.5\%$ and $10.5\%$ more in case of mobilities and scalabilities, respectively (Fig.2.a,d,g,j). Generally, MPRs approach in OLSR provides more optimization in high densities but the conflicting behavior is noticed in Fig. 2.a. Unstable network with high population and dynamicity results MPRs redundancy which expands routing updates dissemination in entire network. Therefore, in VANTEs with more number of connections MPRs fails to provide optimization in the network. By reducing the interval of $LSM$ and $RU\_Tri$ in OLSR-mod (Table. 1), MPRs' are quickly updated, thus makes efficiency better as compared to OLSR-orig in high densities and mobilities (Fig. 2.d,j).

\subsection{E2ED}

In FSR's routing updates are periodic and independent from topological changes as well as degree of nodes, therefore, its routing latency is lower than others (Fig. 2.b,e,h,k). DSR-orig  produces highest E2ED in no. of connections as well as in mobilities due to reactive nature along with RCing and PS. DSR-mod in high mobilities produce high latency due to first checking of RC during ERS augments  the delay in high mobilities, while absence of stale routes and Time-To-Live (TTL) value of $NonPropagating$ RREQ (Table. 1) in larger no. of connections reduces its latency, (Fig. 3.h,k), respectively. OLSR-mod causes highest delay in all scalabilities as compare to other two, because in high densities more number of intermediate hops increase latency.

In FSR-mod, by shortening $RU_{Per}$ interval of inner scope and outer scope results quick updates of routing entries, therefore, delay is reduced as compared to FSR-orig up to $40\%$ (Fig. 2.h,k). In OLSR-mod, routing delay is increased at the cost of throughput, because shorter interval of routing exchange messages makes OLSR-mod more suitable for maintaining accurate MPRs value, while quick detection of link breakage through shorter HELLO interval $RU\_Tri$ for TC messages provides more convergence, therefore achieves high efficiency.

\subsection{NRL}

In DSR-orig, grat. RREPs are generated during RD process which are not suitable in more dynamic or more scalable networks. As, these RREPs produces broadcast storm in case of high dynamic networks through dissemination of incorrect routes because of stale routes in RC. Whereas, in more no. of connections, grat. RREPs from more number of nodes are generated during RD due to congested network. There is no explicit mechanism to delete stale routes except that of limited generation of RERR messages.

In DSR-orig, $1024$ size for $TAP\_CACHE\_SIZE$ stores more faulty routes as compared to DSR-mod which reduce storage of stale routes by decrementing $TAP\_CACHE\_SIZE$ to $256$. This modification in DSR lessens routing overhead up to $-34\%$, as shown in Fig. 2.c,f. Whereas, in high mobilities ($0s$ Pause Time), stale routes are not interrupt to stop ERS, and thus high efficiency of DSR-mod is obtained in term of successful data packet delivery at the cost of more control packets.

Both in FSR-mod and OLSR-mod, shortening update intervals increases number of control messages (in Fig. 2.c,f,i,l). As, there is much difference of $RU$'s exchange interval of FSR-mod (Table. 1) as compared with OLSR-mod which not differs as much as that in FSR (Table. 1). Therefore OLSR-mod augments routing load up to $8.32\%$ compared to OLSR-orig, whereas, FSR-mod produce $10\%$ more control packets as compared to FSR-orig.
 
\section{Conclusion}

We evaluate performance of routing protocols; DSR, FSR and OLSR in VANETs by presenting a framework for $802.11p$ and routing packet cost of these protocols at network layer. Moreover, we have enhance the selected protocols as mentioned in Table. 1. Extensive stimulation work is done in NS-2 for comparison of default versions of three protocols with their enhanced versions. Overall, DSR-mod outperforms rest of the protocols. Enhanced FSR gives more convergence in mobilities by achieving $10.5\%$ high throughput. Whereas, shortening periodic intervals of OLSR makes the MPRs more stable and thus achieves $2.6\%$ more throughput in all scalabilities.

%In future, we are interested to investigate the same study with Linear Programming (LP) modeling for routing in Wireless Multi-hop Networks.

%\ifCLASSOPTIONcaptionsoff
%  \newpage
%\fi

%\vspace{-0.3cm}

  %\subfigure[Different Scalabilities]{\includegraphics[height=3.1 cm,width=5.5 cm]{sca_thru.eps}}
 %\subfigure[Different Mobilities]{\includegraphics[height=3.1 cm,width=5.5 cm]{mob_thru.eps}}
 %\subfigure[Different Scalabilities]{\includegraphics[height=3.1 cm,width=5.5 cm]{sca_e2ed.eps}}
 %\subfigure[Different Mobilities]{\includegraphics[height=3.1 cm,width=5.5 cm]{mob_e2ed.eps}}
 %\subfigure[Different Scalabilities]{\includegraphics[height=3.1 cm,width=5.5 cm]{sca_nrl.eps}}
 %\subfigure[Different Mobilities]{\includegraphics[height=3.1 cm,width=5.5 cm]{mob_nrl.eps}}

\end{document}